\documentclass{emulateapj}
\pdfoutput=1
\usepackage{graphicx}
\usepackage[subrefformat=parens,labelformat=parens,caption=false]{subfig}
\usepackage{url}
\usepackage[normalem]{ulem} 
\usepackage{hyperref}

\shorttitle{Orbital period parameters of Cygnus X-3}
\shortauthors{Bhargava et al.}

\usepackage{xkeyval,xcolor} 	
\makeatletter
\newcommand\footnoteref[1]{\protected@xdef\@thefnmark{\ref{#1}}\@footnotemark}
\newlength{\sfp@hseplen}\newlength{\sfp@vseplen}
\define@cmdkey{subfigpos}[sfp@]{pos}[ul]{}
\define@cmdkey{subfigpos}[sfp@]{font}[\small]{}
\define@cmdkey{subfigpos}[sfp@]{vsep}[2\baselineskip]{\setlength{\sfp@vseplen}{\sfp@vsep}}
\define@cmdkey{subfigpos}[sfp@]{hsep}[10pt]{\setlength{\sfp@hseplen}{\sfp@hsep}}
\newcommand{\subfigimg}[4][,]{%
  \setkeys{Gin,subfigpos}{pos,font,vsep,hsep,#1}
  \setbox1=\hbox{\includegraphics[#3]{#4}}
  \ifnum\pdfstrcmp{\sfp@pos}{ul}=0
    \leavevmode\rlap{\usebox1}
    \rlap{\hspace*{\sfp@hsep}\raisebox{\dimexpr\ht1-\sfp@vsep}{\sfp@font{#2}}}
    \phantom{\usebox1}
  \else\ifnum\pdfstrcmp{\sfp@pos}{ur}=0
    \leavevmode\usebox1
    \llap{\raisebox{\dimexpr\ht1-\sfp@vsep}{\sfp@font{#2}}\hspace*{\sfp@hsep}}
  \else\ifnum\pdfstrcmp{\sfp@pos}{lr}=0
    \leavevmode\usebox1
    \llap{\raisebox{\sfp@vsep}{\sfp@font{#2}}\hspace*{\sfp@hsep}}
  \else
    \leavevmode\rlap{\usebox1}
    \rlap{\hspace*{\sfp@hseplen}\raisebox{\sfp@vsep}{\sfp@font{#2}}}
    \phantom{\usebox1}
  \fi\fi\fi
}
\makeatother

\begin{document}

\title{A precise measurement of the orbital period parameters  of Cygnus X-3}

\author{Yash Bhargava\altaffilmark{1,2},
 A. R. Rao\altaffilmark{1}, 
K. P. Singh\altaffilmark{1},
 Manojendu Choudhury\altaffilmark{3},
 S. Bhattacharyya\altaffilmark{1},
 S. Chandra\altaffilmark{1},
  G.C. Dewangan\altaffilmark{2},
   K. Mukerjee\altaffilmark{1},
    G.C. Stewart\altaffilmark{4},
  D. Bhattacharya\altaffilmark{2},
  N.P.S. Mithun\altaffilmark{5},
S.V. Vadawale\altaffilmark{5},
}

\altaffiltext{1}{Department of Astronomy and Astrophysics, Tata Institute of Fundamental Research, Homi Bhabha Road, Mumbai, India; \email{yash@iucaa.in}}

\altaffiltext{2}{Inter University Center for Astronomy \& Astrophysics, Pune, India}

\altaffiltext{3}{UMDAE Center for Excellence in Basic Sciences, Mumbai University, Mumbai, India}

\altaffiltext{4}{Dept. of Physics \& Astronomy, University of Leicester, Leicester,   UK}

\altaffiltext{5}{Physical Research Laboratory, Ahmedabad, India }
 
\begin{abstract}
We present X-ray light curves of Cygnus X-3 as measured by the recently launched $AstroSat$ satellite. The light curve folded over the binary period of 4.8 hours shows a remarkable stability over the past 45 years and we find that we can use this information to measure the zero point to better than 100 s. We revisit the historical binary phase measurements and examine the stability of the binary period over 45 years. We present a new binary ephemeris with the period and period derivative determined to an accuracy much better than previously reported. We do not find any evidence for a second derivative in the period variation. The precise binary period measurements, however, indicate a hint of short term episodic variations in periods. Interestingly, these short term period variations coincide with the period of enhanced jet activity exhibited by the source. We discuss the implications of these observations on the nature of the binary system.

\end{abstract}

\keywords{accretion, accretion disks — binaries: close — stars: individual (Cygnus X-3) — X-rays: binaries }

\maketitle
\section{Introduction} \label{intro}

Cygnus X-3 (hereafter Cyg~X-3) is an enigmatic X-ray binary. It is bright in almost all parts of the electromagnetic spectrum except the optical band likely due to heavy extinction inherent to the source \citep{Milgrom1976, BonnetBidaud1988, Fender1999}. The binary system consists of a compact object which accretes matter from the companion star. The nature of the compact object of the source is still uncertain due to the absence of reliable mass function estimates. \cite{Zdziarski2013} have presented arguments based on the radial velocity variation of the compact object from X-ray spectral lines and the mass loss rate to conclude that the mass of the compact object is $2.4_{-1.1}^{+2.1} M_{\odot}$, which could either be a neutron star or a black hole. The broadband spectral properties, however, suggest a low mass black hole \citep{Vilhu2009, SzostekZM2008, Szostek2008}. The orbital modulation of the X-ray flux has been explained by a low temperature cloud of plasma surrounding the compact object asymmetrically, which Compton scatters the radiation \citep{Zdziarski2010}.
\par
 
The X-ray light curve of the source has a well documented binary orbital modulation of 4.8 hours. 
The stability of this strong binary modulation of Cyg~X-3 is one of the characteristics that sets this source apart \citep{Ghosh1981}. 

Apsidal motion in the source is not detected in the time frame of $\sim 40$ years \citep{ns2002} thus ruling out any significant eccentricity in the binary orbit \citep{Ghosh1981}. 
The short binary period of 4.8 hours implies a small binary separation and combined with the fact that the companion is a Wolf-Rayet star it shows that the Roche lobe radius of the secondary is indeed tight. The system has undergone $\sim 87000$ revolutions since the first observation and the only long term variation in the period is observed to be secular which indicates that the system is tidally synchronized and phase locked. 

Therefore it is important to confirm the stability of the binary period at longer time scales and in this paper, we provide an improved binary ephemeris. An examination of the evolution of the arrival time is very crucial as it discerns the short-term change in the $\dot{P}$ which demands a physical explanation not hitherto attempted for this source.

 \par
 The evolution of the binary period of Cyg~X-3  has been studied by many authors \citep{Leach1975, Parsignault1976, Manzo1978, Mason1979, Lamb1979, Elsner1980, vdk1981, Kitamoto1987, vdk1989, ns2002}. \cite{Lamb1979} gave an estimate of the period and its derivative while using the data of \cite{Mason1976}. A template for  the binary modulation profile was given by \cite{vdk1989} which was utilized in the current work.

  In this paper, we present the binary folded X-ray light curve of Cyg~X-3, using data obtained from two instruments of $Astrosat$ \citep{Singh2014ASTROSAT, Singh2016} and all available archival data from various instruments. An investigation of the stability of the binary profile for the past 45 years is made, and a precise measurement of the binary ephemeris derived are presented.

\section{Observations and Archival data}\label{sec:data}
\subsection{$AstroSat$ observations}
The Indian multi-wavelength satellite $AstroSat$ was launched on 2015 September 28, and as a part of the Performance Verification phase, Cyg~X-3 was monitored, in particular for a full day on 2015 November 13 and half a day on November 27. $AstroSat$ contains three co-aligned X-ray instruments and an all-sky monitor \citep{Singh2014ASTROSAT}. The data analysis pipeline and the inter-instrumental
cross-calibrations are being streamlined, and in this work, we present only the flux measurement of Cyg~X-3 using the Soft X-ray Telescope and the relative hard X-ray (above 30~keV) binary profile as obtained from the CZT Imager instrument.

\subsubsection{$AstroSat$-SXT observations}

\begin{figure}[htb]

\subfloat{
\subfigimg[width=0.265\paperwidth ,pos=ur,vsep=12pt,hsep=13pt]{(a)}{angle=-90}{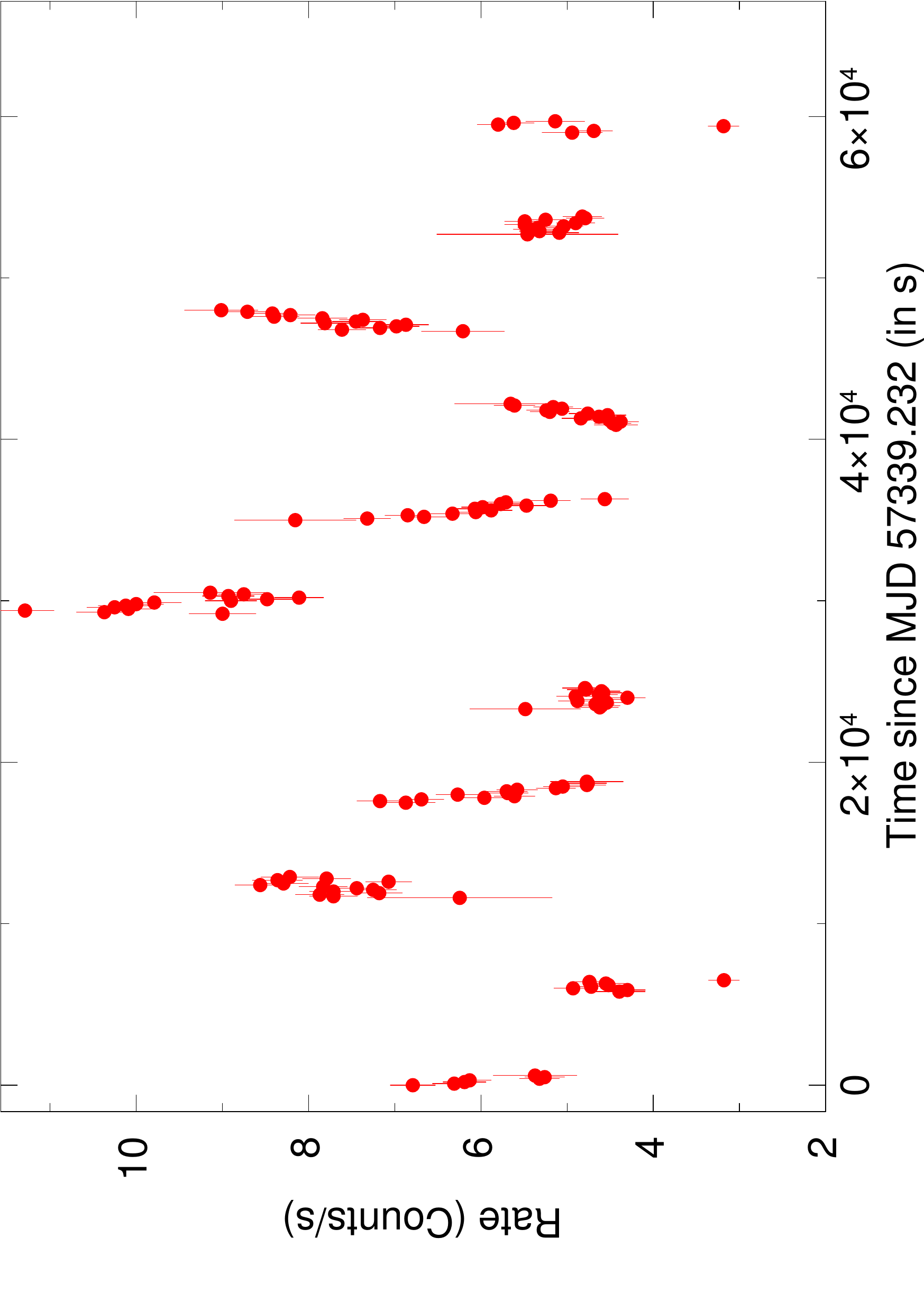}
\label{fig:lc_sxt_a}
}

\subfloat{

\subfigimg[width=0.265\paperwidth ,pos=ur,vsep=12pt,hsep=13pt]{(b)}{angle=-90}{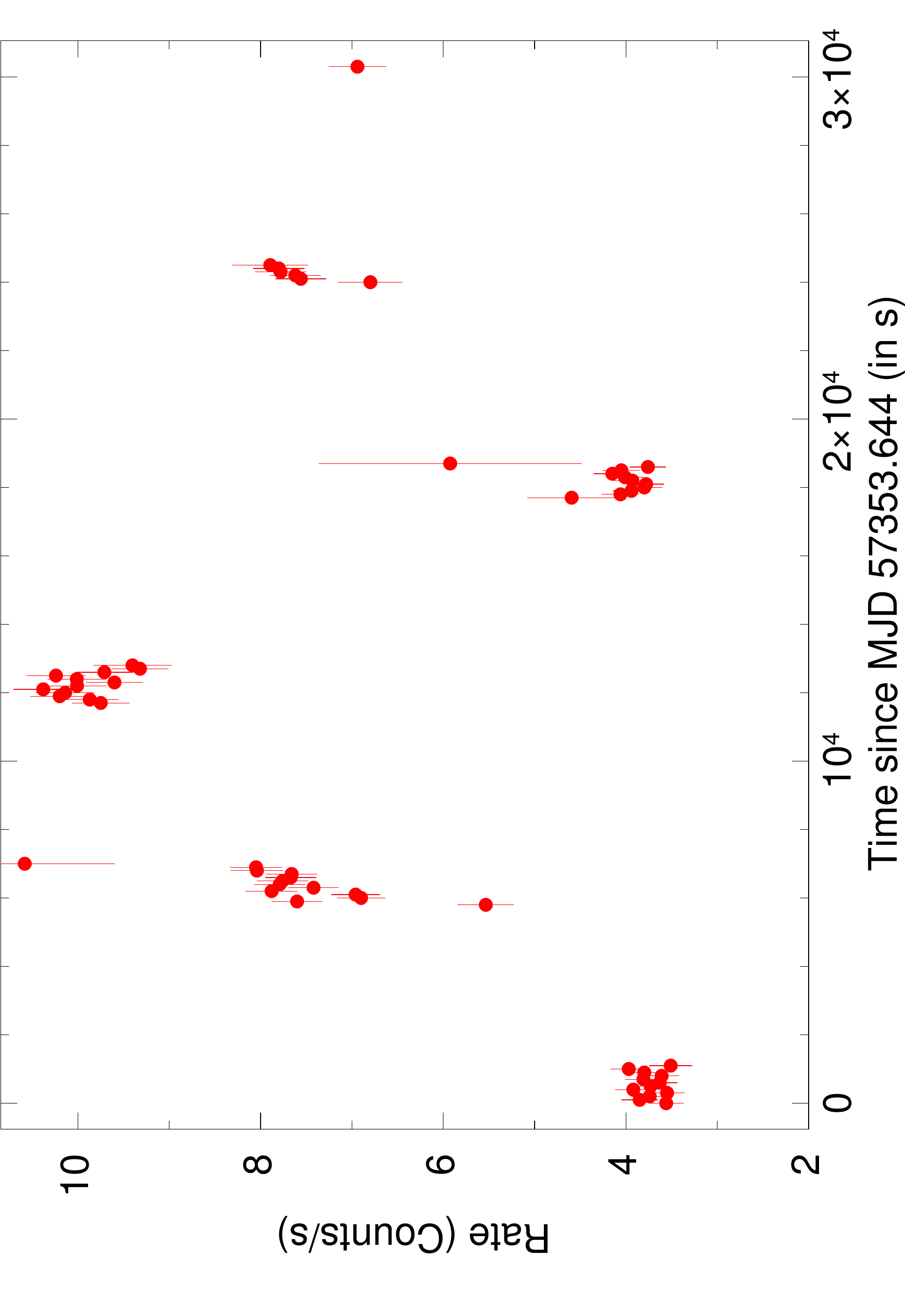}
\label{fig:lc_sxt_b}
}
\caption{Lightcurves obtained from SXT observations of Cyg~X-3 made on 2015 November 13 (a) and 27 (b) in the energy range 0.7--7 keV with a bin size of 100~s. The variation in counts due to the binary modulation can be clearly seen. The gaps in the data set are due the bright earth limb and Earth occultation of the source.\footnote{The data for these figures can be downloaded from  \href{http://astrosat-ssc.iucaa.in/uploads/sxt/SXT\_LightCurves.zip}{http://astrosat-ssc.iucaa.in/uploads/sxt/SXT\_LightCurves.zip}}} 
\label{fig:lc_sxt}
\end{figure}

The Soft X-ray Telescope (SXT) aboard $AstroSat$ is a grazing incidence X-ray telescope with the focal length of 2~m  \citep{Singh2014ASTROSAT, Singh2016}. SXT observed Cyg~X-3 on 2015 November 13 and  27 in Photon Counting (PC) mode. 
The detector used in the focal plane of the telescope is a Charged Coupled Device, CCD-22 of e2V Technologies Inc., UK. The effective area of the SXT is  $\sim $128 cm$^2$ at 1.5 keV. The energy range of the instrument is 0.3 -- 8~keV, with an energy resolution of 5-6\% at 1.5~keV. The field of view is $ \sim 40'$ dia and the point spread function can be described by a double King function with FWHM of $ \sim 40$\arcsec\ for the inner core (see \cite{Singh2016}). $AstroSat$ is in a low earth orbit at an altitude of 650 km with an inclination of 6$^\circ$  and an orbital period of 98 min. 
The Earth occultation, the passages through South Atlantic Anomaly (SAA) region and the sensitivity of the detector to the bright limb of the Earth reduces the useful data to $\sim$2 ks every orbit. The  light curves with a bin size of 100 s and 0.7--7 keV energy range were extracted using a circular extraction region of 15 arcmin radius from observations on November 13 and 27 and are shown in Figure \ref{fig:lc_sxt}\subref{fig:lc_sxt_a} and \subref{fig:lc_sxt_b} respectively. \par

\begin{figure}[!htb]
\subfloat{
\subfigimg[width=0.27\paperwidth ,pos=ur,vsep=12pt,hsep=12pt]{(a)}{angle=-90}{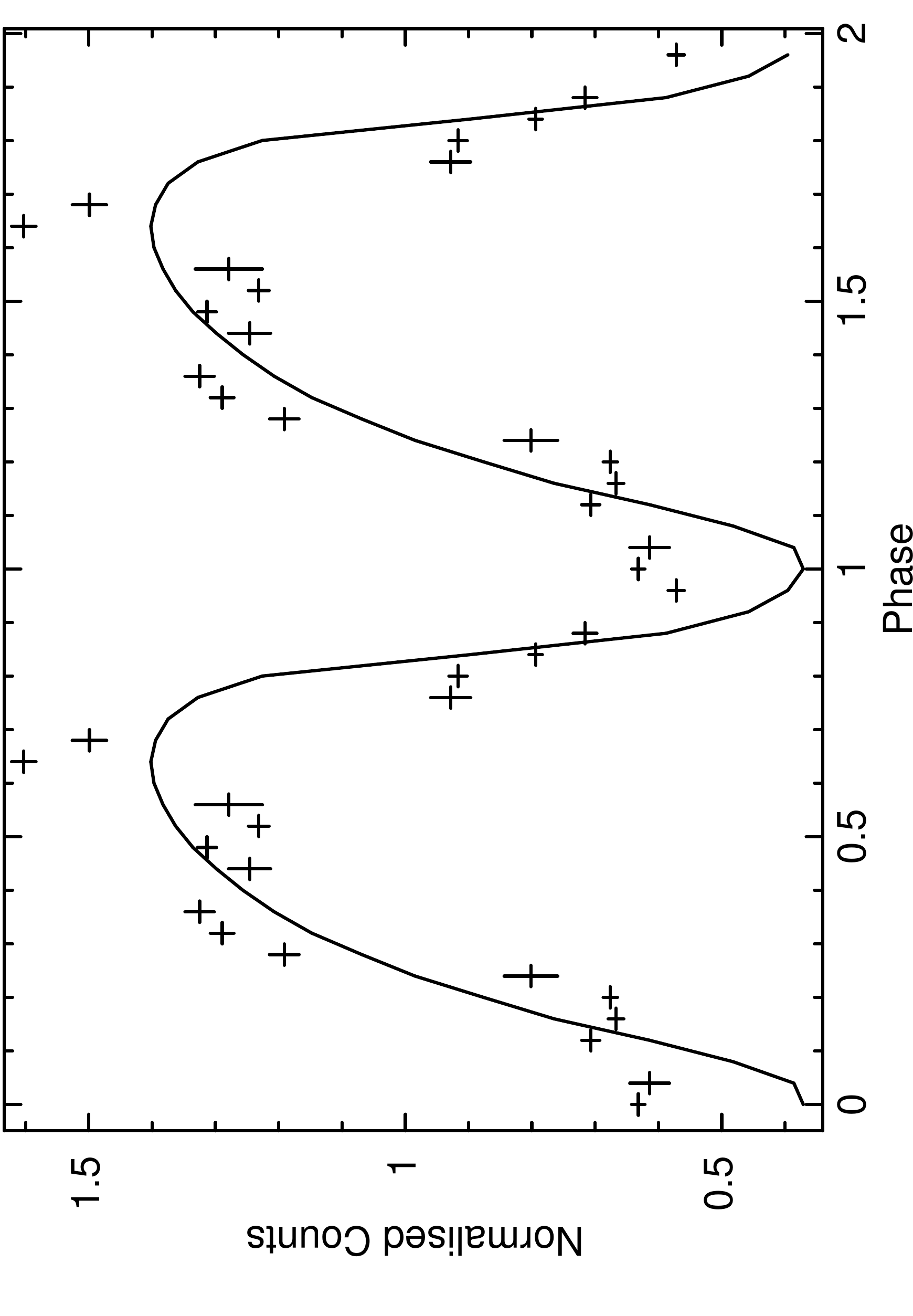}

\label{fig:fold_sxt}
}

\subfloat{

\subfigimg[width=0.27\paperwidth ,pos=ur,vsep=12pt,hsep=9pt]{(b)}{angle=-90}{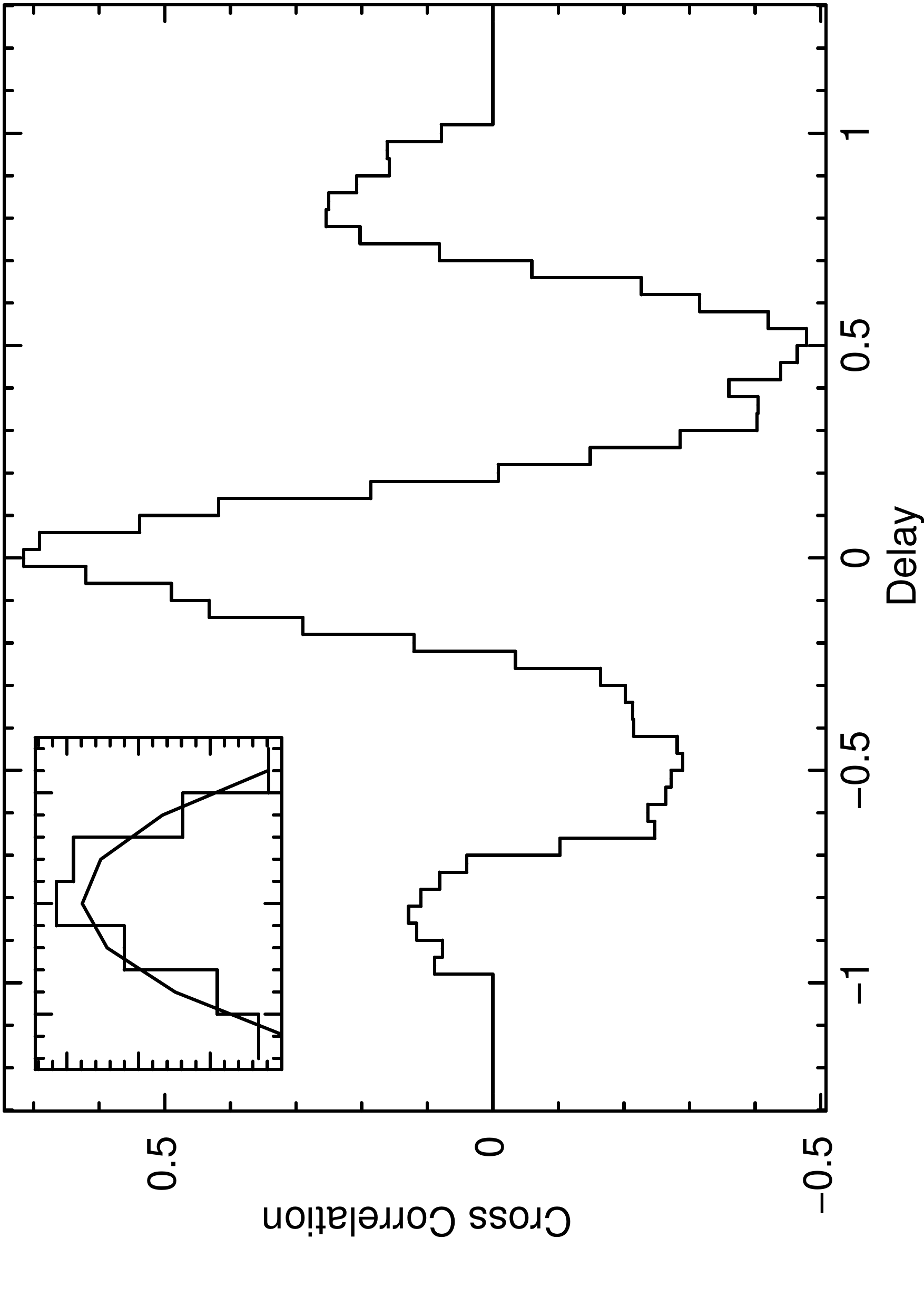}
\label{fig:cross_sxt}
}

\caption{\emph{(a)} The SXT data of Cyg~X-3  folded onto the binary ephemeris as given by \cite{ns2002}, overlaid with the binary template \citep{vdk1989}.  \emph{(b)}  Cross-correlation function between the template and the folded light curve. A quadratic fit to the cross-correlation function is shown in the inset. The zero phase is measured to be shifted by 0.0029$\pm$0.001 (50.7$\pm$17.0 s) with respect to the expected zero phase based on the ephemeris.}
\label{fig:sxt_analysis}
\end{figure}

The lightcurves were folded using the ephemeris from \cite{ns2002}. The data were folded into 25 phase bins. The folded data are shown in Figure \ref{fig:sxt_analysis}\subref{fig:fold_sxt} and the binary template, as given by \cite{vdk1989}, is overplotted as a curve in the figure. The folded data were cross-correlated with the template to determine the deviation in the arrival time. We measured the zero phase, with respect to the template shifted by the ephemeris, to be 0.0029$\pm$0.001 (50.7$\pm$17.0 s). The details of the measurement have been discussed in Section \ref{sec:cross}.

\subsubsection{$AstroSat$-CZTI observations}

\begin{figure}[htb]
	\centering
	\includegraphics[width=0.45\paperwidth]{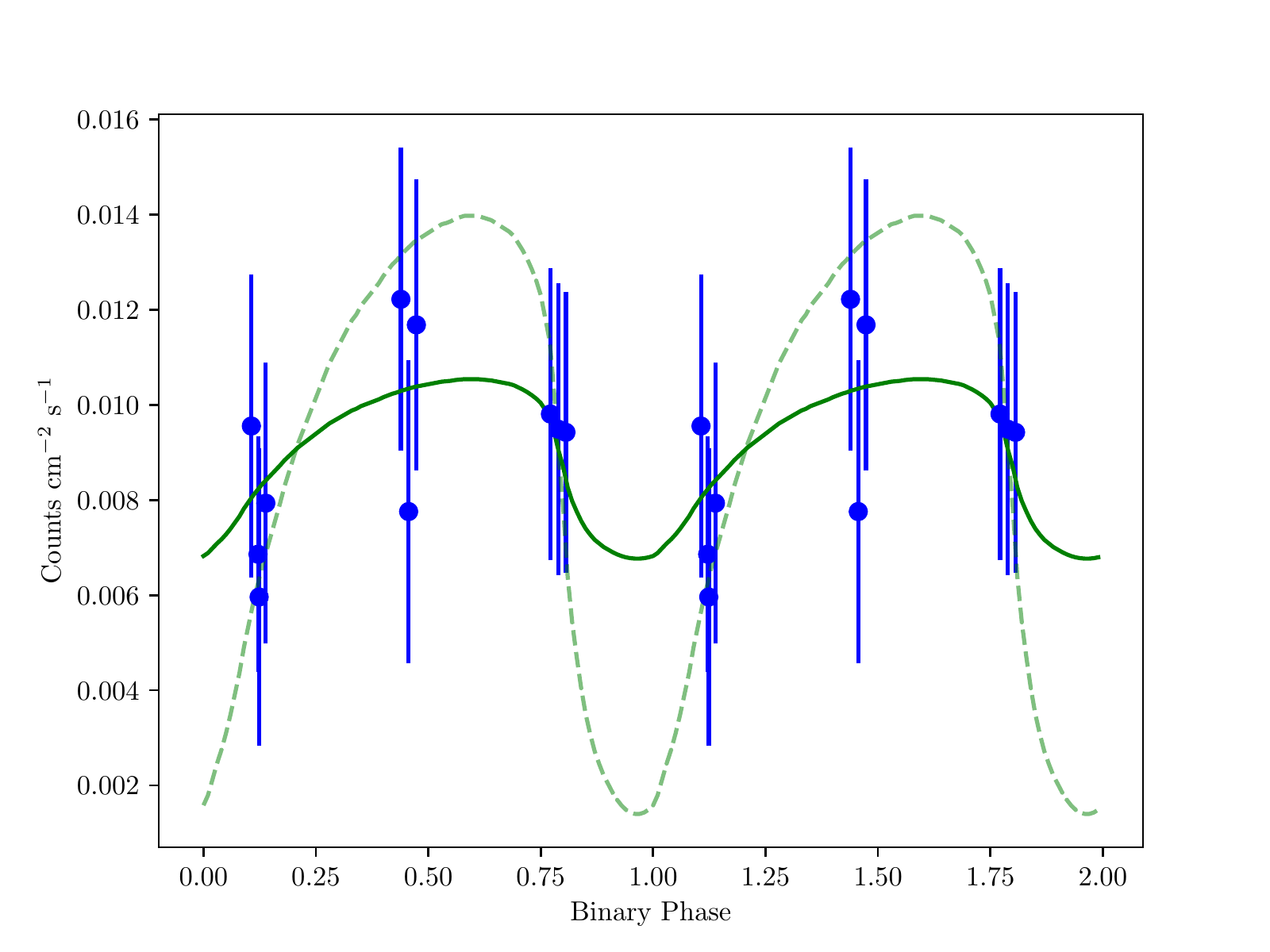}    
	
	\caption{Hard X-ray ($>$30 keV) observations of Cyg~X-3 using $AstroSat$ CZTI, folded onto the binary period. The background subtraction for the instrument was done by using mask weighting technique \citep{Bhalerao2016}. The counts were thus computed using a average bin size of 1000 s. The binary template of Cyg~X-3, scaled to a modulation depth of 30\%,  is shown as a continuous line. The template with 100\% modulation depth is also shown, for comparison, as a dashed line. }
	\label{fig:czti}
\end{figure}

Cyg~X-3 shows binary modulation in hard X-rays (above 10 keV), though at a reduced modulation depth \citep{Rajeev1994}.  We have analyzed the hard X-ray (above 30 keV) data obtained from the $Astrosat$ CZTI instrument. The  CZTI instrument is sensitive in the 20--200~keV range (though for the observations presented here a low energy threshold of 30~keV was used) and uses  Cadmium Zinc Telluride Detectors and a coded aperture mask  \citep{Bhalerao2016}.  CZTI is co-aligned with the SXT instrument and observations are made contemporaneous with the SXT observations. We have used a mask weighting method to obtain the hard X-ray flux from  Cyg~X-3, and the resultant data are shown in Figure \ref{fig:czti}. The Crab Nebula gives around 0.14 counts cm$^{-2}$ s$^{-1}$  in CZTI and we deduce that the average flux of Cyg~X-3 during the present observations is about 60~milliCrab.  We plot the binary template in Figure \ref{fig:czti} scaled to a modulation depth (ratio of peak-to-peak variation to average) of  30\% (solid line in the figure), in contrast to the $\sim$100\% modulation seen in the low energy SXT data (dashed line in the figure).  Though the data do not show significant modulation, they  are consistent with the low modulation seen at high energies  \citep{Zdziarski2012}.

\subsection{Archival Data}
\cite{ns2002} have presented the binary ephemeris of Cyg~X-3 based on   \emph{IXAE}-PPC data obtained in 1999,  \emph{RXTE}-ASM data up to 2001, and other archival pointed observations in 1994 -- 1998 period along with the binary arrival times reported in the literature. To connect to the recent \emph{AstroSat} data, we have analyzed the complete archival data from the \emph{RXTE}-ASM (till 2012). To perform  a uniform analysis, we have also re-analysed the archival data from the \emph{Einstein} observatory \citep{Elsner1980},  the \emph{EXOSAT} \citep{vdk1989},  and the \emph{Ginga-LAC}  \citep{Turner1989}, obtained from HEASARC archives. The  \emph{Einstein Observatory} had seven pointed observations and data from the Monitor Proportional Counter (MPC) ($1-20$ keV) were taken. Due to the short duration of individual files, the files were combined into two sets, corresponding to different years of observation. Data from \emph{EXOSAT}'s Medium Energy (ME) instrument ($1-50$ keV)  were used for obtaining the arrival points. \emph{EXOSAT} had 5 data segments which were greater than the modulation period and were folded individually. Rest of the data were combined and folded to generate another data segment. Light curves from Large Area Proportional Counters (LAC) onboard \emph{Ginga}  ($1.5-37$ keV) were used to get the arrival points during that period. The data from LAC were merged to generate folded light curves, as individual files had low counts. Table \ref{tab:data_det} lists the times and durations of the observations and the data points extracted from these missions.

The All Sky Monitor (ASM) onboard the \emph{RXTE} satellite \citep{Levine1996} had continuous monitoring of the source in energy range $1.3 - 12$ keV which helps in tracking the number of orbits of the source.ASM observed a large field of view (3 Scanning Shadow Cameras; 6$ ^{\circ} $ by 90$ ^{\circ} $ for each camera) for 90 s and then rotated to cover 80\% of the sky in 90 min. Due to  the sparse nature of the data, larger segments of the data from ASM were taken.Monitor of All-sky X-ray Image (MAXI) onboard the International Space Station also monitors the source but its data was not included in the current work. The available data from the satellite is averaged over the satellite orbit, which has larger inherent error bars.

\begin{table}[!htb]
    \centering
    \caption{Archival data used in analysis}
    \label{tab:data_det}
    
    \scriptsize
    \begin{tabular} 		 
    {ccccc}
        \hline \hline
        \textbf{Satellite} & \textbf{Detector} & \textbf{Start MJD} & \textbf{Stop MJD} & \textbf{Data points} \\ \hline \hline
        \emph{Einstein}    & MPC &    43844.430     & 44004.596     & 2 \\ 
        \emph{EXOSAT}      & ME  &    45518.891     & 46364.799     & 6 \\ 
        \emph{Ginga}       & LAC &    47034.661     & 48125.788     & 2 \\ 
        \emph{RXTE}        & ASM &    50087.323     & 55915.926     & 77 \\  \hline \hline
\end{tabular}
\end{table}
\section{Data Analysis and Results}
\subsection{Binary Phase Analysis}\label{sec:proc}

Light curves from the ASM data were folded at a nearby epoch. Since the period of the source is not constant \citep{ns2002, Kitamoto1995}, the length of the folded data was kept as small as possible. \cite{ns2002} had folded the ASM light curves using a constant period and calculated the epoch point using parameters from \cite{vdk1989}. In the current analysis, however, the number of elapsed binary periods, $n$,   has been calculated using parameters from \cite{ns2002} and the folding period, $P_n$ was computed as

\begin{equation}
\label{eq:folding_period}
P_n = P_0 + \dot{P}P_0(n+\frac{1}{2})
\end{equation}
where $P_0$  and $\dot{P}$ are the period and period derivative as given in  \cite{ns2002}. 
To maintain consistency,  the epoch and folding period were determined in a similar fashion for all datasets. 
For the ASM, the length segment of the datasets was selected based on the count rate.
For the high count rate period, the length segment consisted of 50 days whereas for the low count rate period it was of 100 days. Near the end of the mission, the ASM had even sparser monitoring, and thus the length segment of the datasets was taken to be of 200 days. Longer datasets contained more orbits of the source, and to keep the epoch points sufficiently apart; the epoch was taken as the start of the orbit containing the midpoint of the file. The  epoch $T_n$ is  calculated  using the orbit number $n$ which contains the midpoint of the data segment as 
\begin{equation}
\label{eq:arr_time}
    T_n = T_0 + P_0n + \frac{1}{2}\dot{P}P_0n^2 
\end{equation} \par
where $T_0$ is the epoch as given in  \cite{ns2002}.

 \par

\subsection{Cross correlation and fitting}\label{sec:cross}

\begin{figure}[!htb]
  
    \subfloat{
    \subfigimg[width=0.27\paperwidth,pos=ur,vsep=14pt,hsep=12pt]{(a)}{angle=-90}{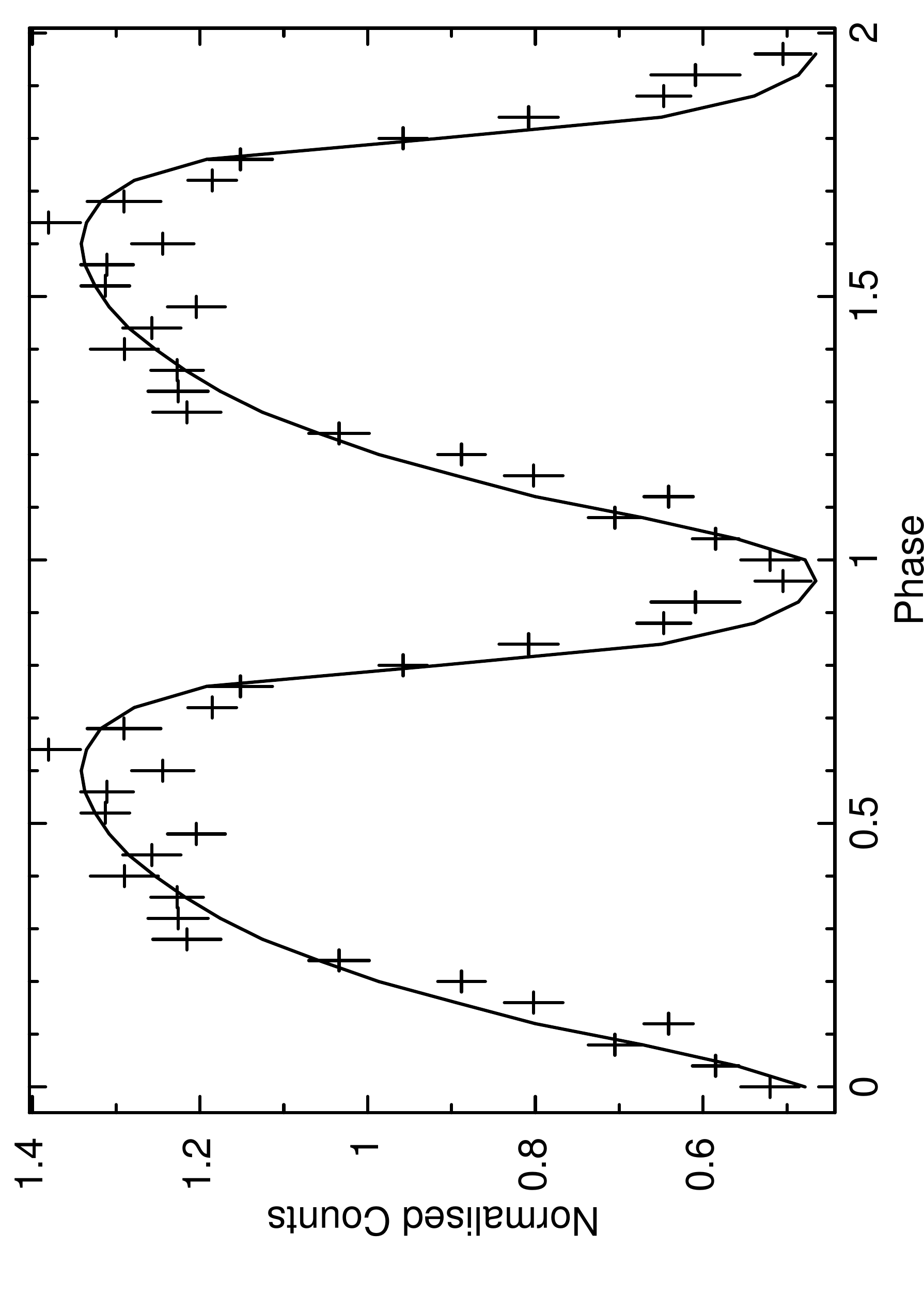}
    \label{fig:pp}}
    
    	\subfloat{
    \subfigimg[width=0.27\paperwidth,pos=ur,vsep=12pt,hsep=9pt]{(b)}{angle=-90}{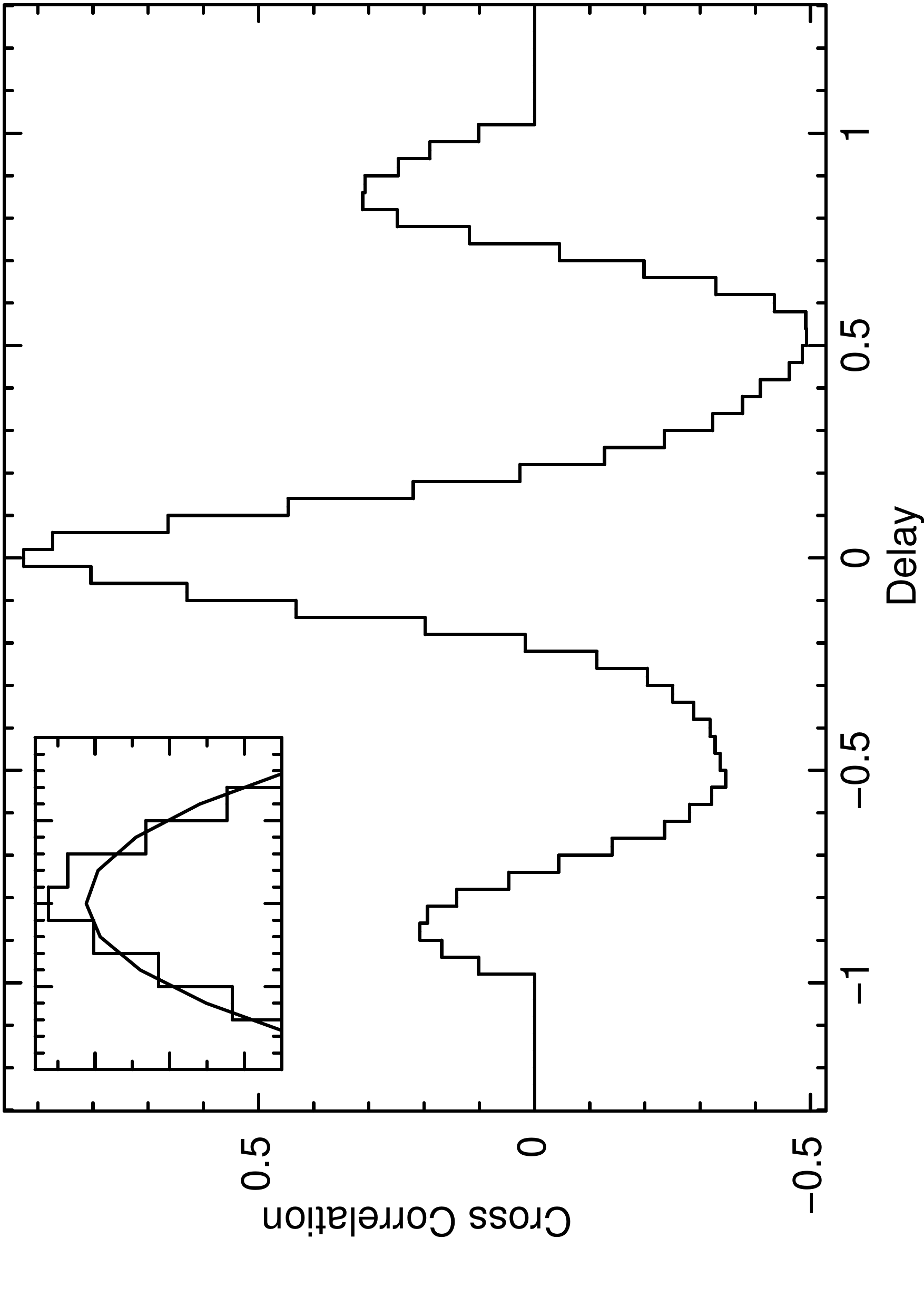}
	\label{fig:cc}}
		
	\caption{ (a) The folded profile of the Cyg~X-3 using \emph{RXTE}-ASM data showing the distinctive feature of a slow rise and small plateau and a steep decline at the end. The folded profile was cross-correlated with the template function\citep{vdk1989} to determine the deviation from the assumed epoch. (b) Cross-correlation function. To determine the actual delay, a quadratic function was fitted as shown in the inset. }
	\label{fig:asm}
\end{figure}

  Folding of the light curves clearly depicts the binary modulation of the source. The modulation can be seen in Figure \ref{fig:sxt_analysis}\subref{fig:fold_sxt} for SXT data and in Figure \ref{fig:asm}\subref{fig:pp} for ASM data. 
To estimate a more precise value of the arrival time, the folded light curves were cross-correlated with the template function from \cite{vdk1989}. In \cite{ns2002} the phase correction was determined by taking the bin corresponding to the maximum cross-correlation coefficient. In the current analysis, we found that the cross correlation function need not be symmetrical about the central bin but symmetrical about the central value. Figures \ref{fig:sxt_analysis}\subref{fig:cross_sxt} and \ref{fig:asm}\subref{fig:cc} shows cross-correlation function which shows the mentioned scenario. To determine the exact position of the peak, a quadratic function was fitted to the points around the peak. The peak bin and three bins on either side of the peak were used for fitting the quadratic function. This method yields a more accurate determination of the peak position. 
\par 
\par

\begin{figure*}[!htbp]
    \centering
	\includegraphics[width=0.49\textwidth]{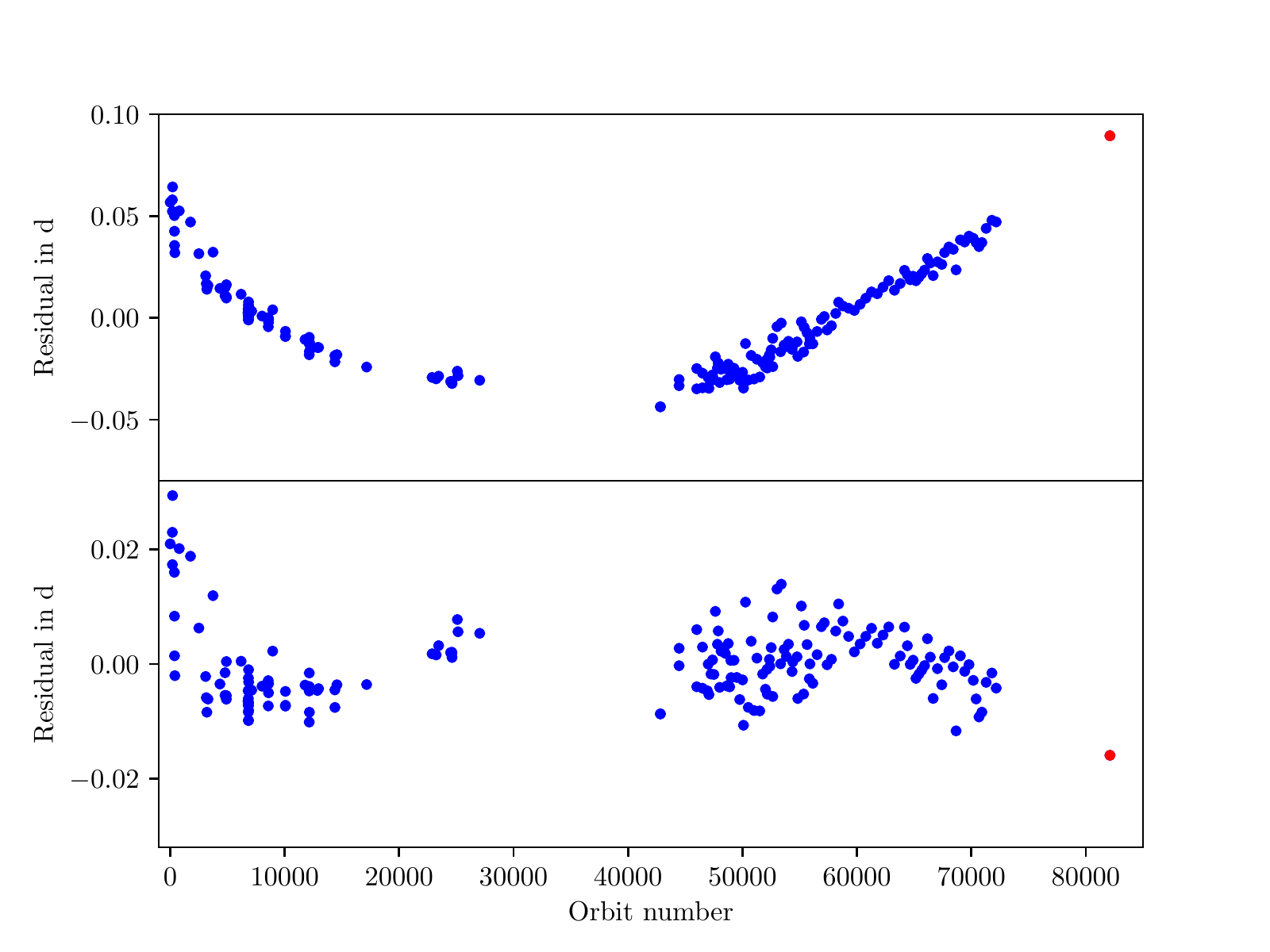}		\includegraphics[width=0.49\textwidth]{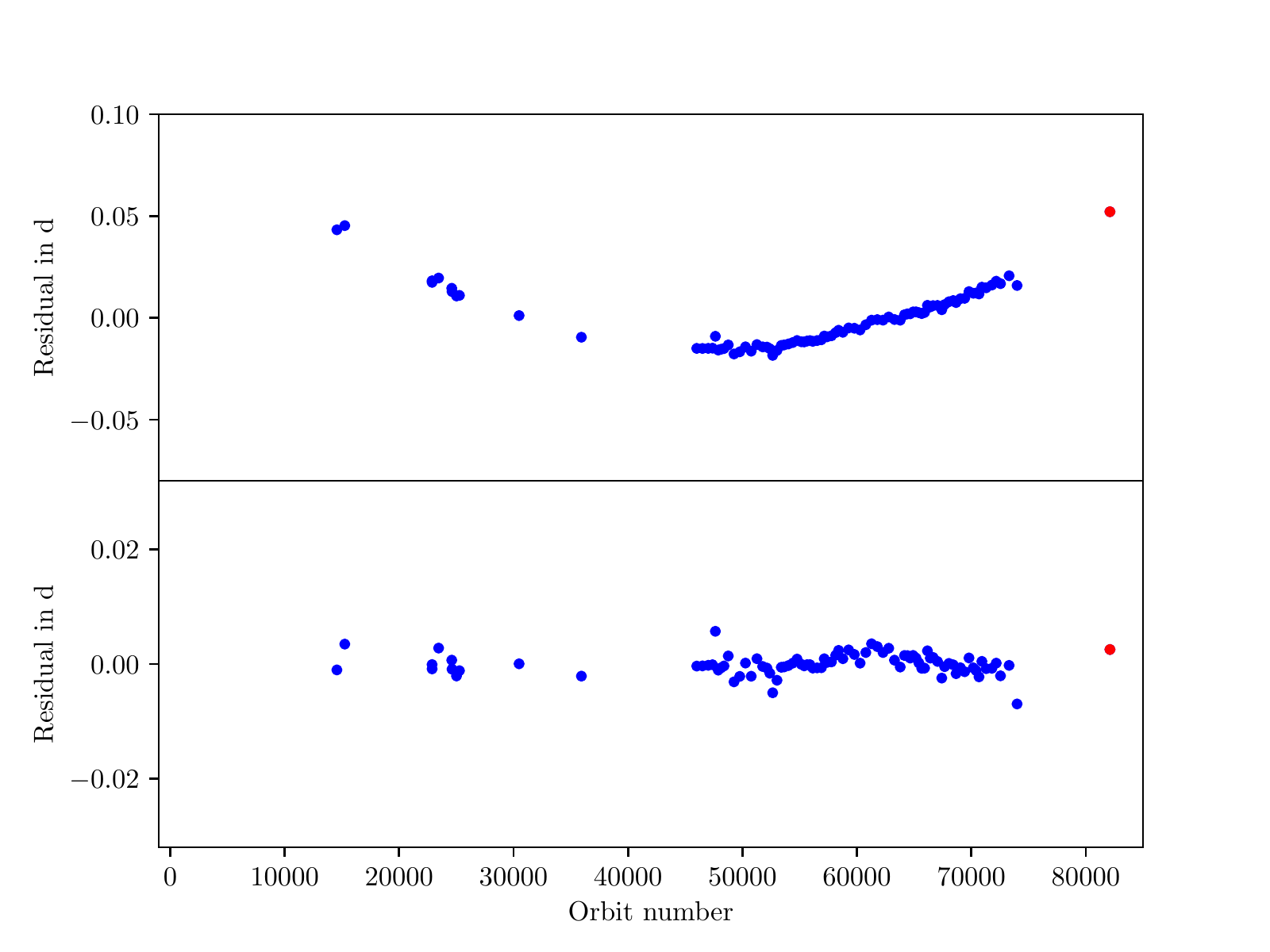} 
    \caption{Residuals from a quadratic fit to the arrival time of Cyg~X-3. In the panels on the left, the arrival points used for computing the residuals have been evaluated using the parameters and method used by \cite{ns2002}. Data beyond orbit 55848 are computed in the present work. Top left and Bottom left panels show the residuals of best-fit linear and quadratic models respectively. In the Right side panels, the arrival points are computed using a method involving period and period derivative for folding and using a quadratic fit to the cross-correlation function (see text). As compared to left panels, the inclusion of a period derivative term in the determination of epoch and fitting the cross-correlation function by a quadratic component leads to a noticeable reduction in the residuals. The red point corresponds to the Arrival time measured by $AstroSat$-SXT.}
    \label{fig:res_prev}

	\label{fig:res_new}
\end{figure*}

The estimated epoch was corrected by the difference between the fitted peak and the estimated epoch to yield the arrival time of the minima. The error on each arrival time was determined using the error bars of the quadratic fitting. Since we measure the slight shift in the phase by fitting a parabola, the error in such measurement is dictated by the error in the fitting.  Following the work of \cite{ns2002} the arrival time and orbit number were fitted with a  linear and a  quadratic function.  Two methods of fitting were undertaken. In the first method, the archival data from  \cite{ns2002} and the arrival points from ASM and SXT computed using the method described in \cite{ns2002} were fitted.  Residuals thus obtained are shown in Figure \ref{fig:res_prev}. In the second method, the data described in Section \ref{sec:data} were fitted. The error bars were assigned by computing the error in peak determination and no further systematic errors were added. The residuals from the second method are shown in figure \ref{fig:res_new}. 
Data fitted using the first method have larger residuals as compared to the ones obtained from the second method. The results obtained from fitting using both the methods are tabulated in Table \ref{tab:Results}. \par

\begin{figure}[!htb]
	\centering
	\includegraphics[width=0.45\textwidth]{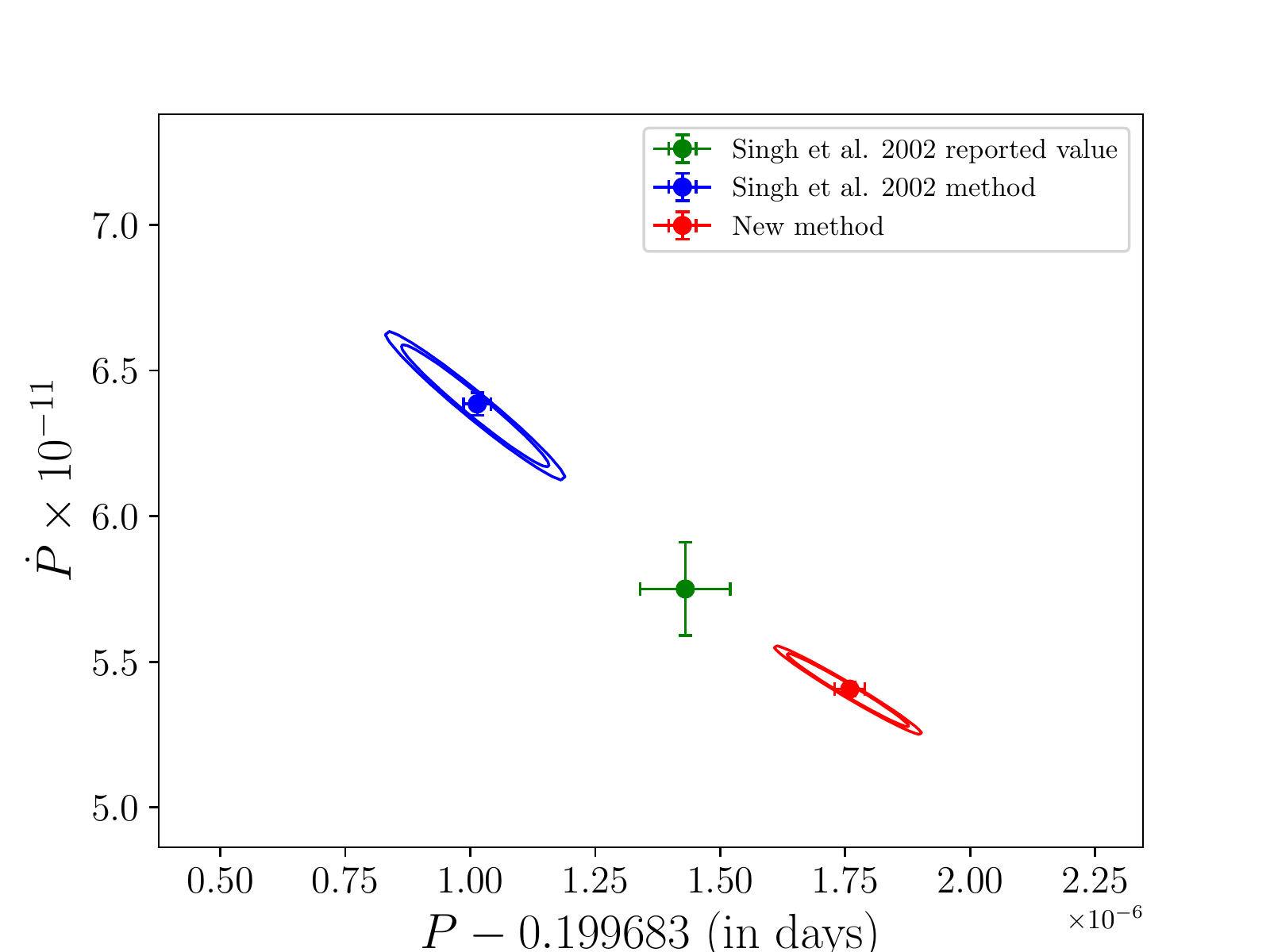}
	\caption{$ \chi^2 $ map giving the estimated values of the parameters along with the confidence ranges on the estimates corresponding to $ \Delta\chi^2 $ of 2.71 and 4 (nominal 90\% and 95\%, respectively) are shown. The parameter values reported by \cite{ns2002} is shown in green at the center. The blue point (left top) corresponds to the parameter obtained by using the method given by \cite{ns2002} (Method 1). The red point (bottom right) similarly corresponds to the method described in the present work (Method 2).} 
	\label{fig:chi2_map}
\end{figure}

The $ \chi^2 $ maps for the P and $\dot{P}$  derived based on the two methods are plotted in   Figure \ref{fig:chi2_map}. The results reported by \cite{ns2002} is shown in the same figure as a plus sign (in green) in the middle of the figure. The contour levels depicted correspond to nominal 90 and 95 \% confidence  ($ \Delta\chi^2 $ of 2.71  and 4  respectively)  for both the cases.  The contour levels shown in blue at left top correspond to the method by \cite{ns2002}, and the red contour levels correspond to the method described in sections \ref{sec:proc} and \ref{sec:cross}. 
 Including a cubic term to fit the residuals (so that the effects of a second derivative to the period are included) resulted in a reduction of $ \chi^2 $ of 41.2 and 2.94 for \cite{ns2002} and the current method respectively while including a sinusoidal term reduced the $ \chi^2 $ by 16.7 and 2.78 respectively. 
 
\cite{Zdziarski2012} smoothens the folded light curves from ASM by computing the running average of the light curves. The authors renormalised the light curves using the average value to remove the long term aperiodic variability. The procedure helps the authors in determining the shape of the modulation with improved precision. The method was attempted for the current analysis and the improvement in the shape was seen without a significant shift in the position of the minima. Therefore the treatment followed by \cite{Zdziarski2012} is not followed here.
 
 It is interesting to note that the inclusion of data points beyond those used by \cite{ns2002}, \emph{RXTE}-ASM data after  2001 and SXT data, indicates a  slowing down of the system (lower period) beyond 2001.  Method 1 gives smaller errors for the derived parameters of P and $\dot{P}$   due to the longer time base of measurement but  gives lower values of the period and higher period derivative. 

 The residuals, however, indicate that in recent times there is an indication of a lower period which is in contradiction with the early longer periods and cannot be compensated with a single period derivative.  
Our analysis (Method 2) that treats all available data uniformly shows that all data beyond orbit 12000 can be fit with a quadratic function, without requiring any systematic errors. The
errors in individual data are much better and, most interestingly, the data do not need any third derivative or periodic components ($\Delta  \chi^2 $ of  2.94 and 2.78, respectively). Hence, we can conclude that beyond orbit number 12000 the Cyg~X-3 is showing only a secular variation (with higher period and lower period derivative than reported earlier). It is possible that the data points before orbit number 12000 have higher systematic errors or the
source showed a short term  episodic variation. 
\par

\par

\begin{table*}[!htbp]
    \centering
    \caption{Comparison of Ephemeris of Cyg~X-3}
    \label{tab:Results}
    \begin{tabular}{|l|l|}
        
        \hline
		
        \multicolumn{2}{|l|}{Quadratic Ephemeris:}   \\
        \multicolumn{2}{|l|}{$ T_n = T_0 + P_0n + cn^2 $}             \\
        \multicolumn{2}{|l|}{where $ c = (P_0\times \dot{P})/2$, $T_{0}$ in MJD-40000}       \\
        \hline
        \cite{ns2002} method					& Present Work \\ \hline
		${\chi}^2 = 182.9$ for 175 dof			& ${\chi}^2 = 77.6$ for 86 dof			\\        
        $T_0 = 949.399 \pm 0.001 $               & $T_0 = 949.384 \pm 0.001 $				\\
        $P_0 = 0.19968401 \pm 0.00000002     $   & $P_0 = 0.19968476 \pm 0.00000003     $		\\
        $c = (6.38 \pm 0.04)\times10^{-11}  $   	& $c = (5.41 \pm 0.02)\times10^{-11}  $   \\
        $\dot{P} = (6.38 \pm 0.05) \times 10^{-10} $ & $\dot{P} = (5.42 \pm 0.02) \times 10^{-10} $  \\
        \hline
        
    \end{tabular}
\end{table*} \par

\begin{figure*}[!htb] 
    \centering
   
       \includegraphics[width=0.8\paperwidth,angle=0]{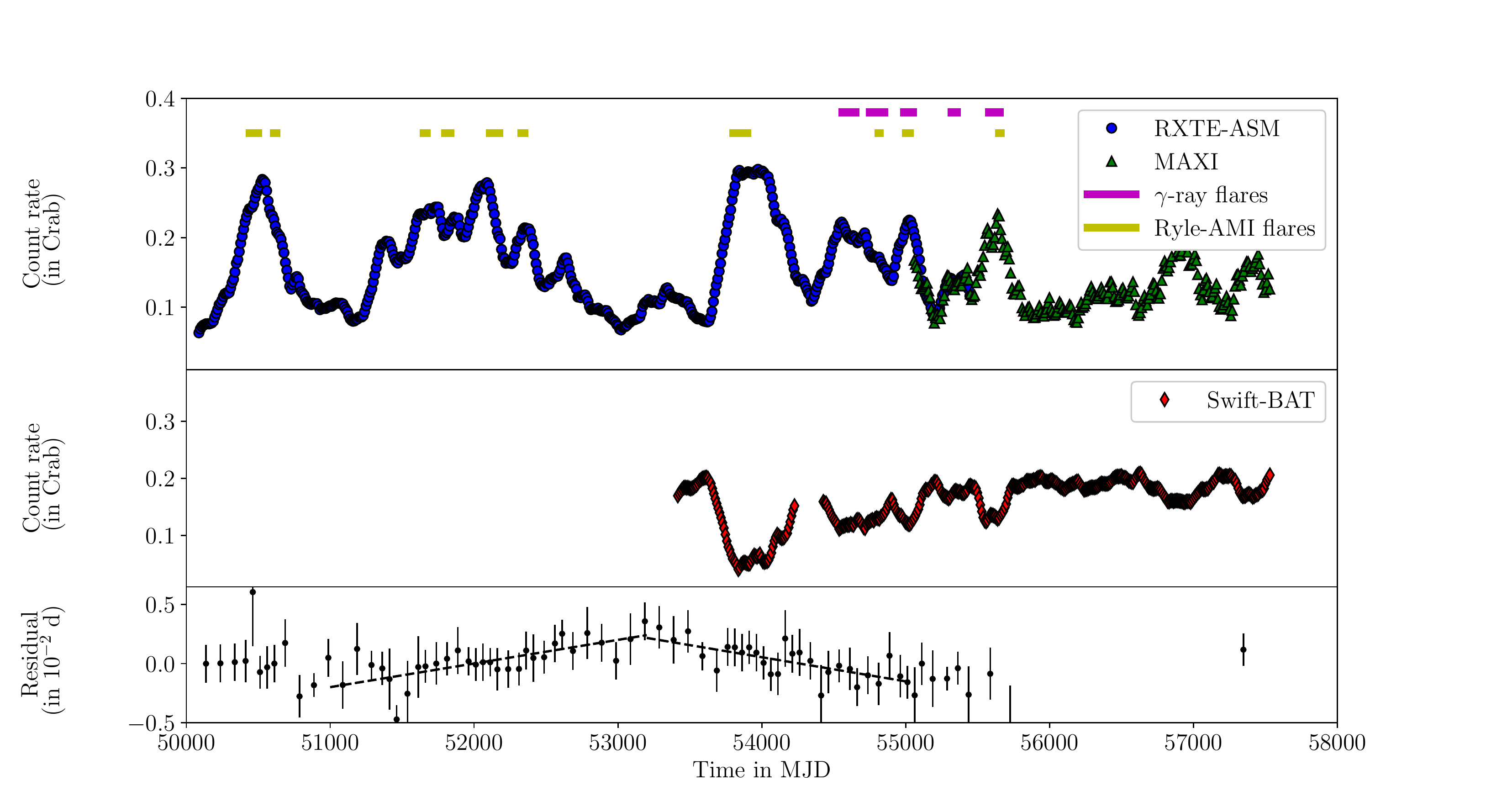}
	
    \caption{X-ray Light curve of Cyg~X-3, binned at 10 days and smoothened by taking a running average for 100 days. Light curves are plotted for the mission duration of ASM (1.3-12 keV; blue circle), \emph{Swift}-BAT(15-150 keV; red diamond) and MAXI(2-20 keV; green triangle). Black dots correspond to the residuals of the fitting function in Table \ref{tab:Results}. Magenta bars correspond to $\gamma$-ray emission from the source while yellow bars correspond to Radio flares detected by Ryle/AMI in 15 GHz, which are greater than 100 mJy in flux, as given in  \cite{Zdz2016}. The light curve has been normalized to Crab counts. Binary modulation cannot be seen, as it is lost in the binning of the data. The residuals show a broad feature indicating a change in the period from MJD 51000 to 55000. Straight line fits to the rise and fall part of this feature are shown as continuous lines in the panel
(see text for details).      }
    \label{fig:lc10all}
\end{figure*}

The residuals from the ASM data show a broad hump which indicates a possible change in parameters over large orbit numbers. To examine these changes and to investigate their association with the other properties of the source,    the residuals are plotted along with  the heavily smoothed light curve of the source in  Figure \ref{fig:lc10all}. The light curve was smoothed by binning by  10 days with a running average of 100 days. The light curves from the \emph{Swift} - Burst Alert Telescope (BAT), and the Monitor for All-sky X-ray Image (MAXI) were also included to indicate the behavior of the source over the years. The light curves from different instruments have been normalized to Crab. Prominent radio ($>$ 100mJy from Ryle/AMI) and $\gamma$-ray flares were also included in the figure, as given in  \cite{Zdz2016},  to qualitatively correlate the variation in the residuals to source activity. 

A  broad hump in the timing residuals is seen during the period  MJD 51000 to 55000. To understand the 
significance of this variation, we have further subdivided this period into 'rise' and 'fall' regions. The rise corresponds to MJD 51000 to  53200 and the fall to MJD 53200 to 55000. The rise and fall regions  were independently fitted with a constant and a straight line. A F-test showed that the   f-ratio   for the rise is 22.23 and p-value is 6.03$\times10^{-5}$ and for the fall, f-ratio is 24.71 and p-value is 4.46$\times10^{-5}$. 
The significance, however, would be diluted by the number of trials (we conservatively estimate it to be about  10)  because the selection of the rise and fall periods are subjective. Nevertheless,   we can conclude that there 
is an indication of  a statistically significant feature in the period MJD 51000 to 55000, which invites the discussion over it's physical implication. We also note that over the complete duration of analysis for period variation (MJD 50000--58000), this short term variation does not have any significant impact on the results.  A F-test, as above, for the same yielded a f-ratio of 0.2 and p-value of 0.65 for any additional linear term, leading to the conclusion that a simple model is a better explanation of the residuals in the full range. 

\par

\section{Discussion}

Cyg~X-3 is a remarkable X-ray binary showing emission from radio to high-energy gamma-rays, but it defies a consistent explanation based on our current understanding of black hole binaries.  Though it shows spectral and multi-wavelength variability behavior similar to those seen in other black hole binaries like Cyg X-1 and GRS 1915+105 \citep{Manoj2002, Manoj2003, Manoj2004} each harboring a black hole of mass  $\sim$14~M$_\odot$, the available data on Cyg~X-3 is consistent with a low mass black hole \citep{Zdziarski2013}.  Compared to Cyg~X-1 and GRS~1915+105, Cyg~X-3 has a much lower binary period. 
By making a systematic analysis of the secular variation of the binary period, we have demonstrated here that the data are consistent with a simple secular variation of the period, without the requirement of any second derivative or periodic variations. There are, however, indications of small local period variations (e.g. MJD 50000-58000) possibly associated with the jet emission properties.

Strong winds from the companion Wolf-Rayet star are the obvious reason for the secular evolution of the periodicity of this source. At a coarser level the total mass loss in a binary system may be related to the secular increase of periodicity by the relation $\dot{M}_{tot}=-M_{tot}\dot{P}/2P$ where $M_{tot}$ is the total mass of the binary stars \citep[see, for eg.][]{Kitamoto1995}. Given the ambiguity of the nature of the Wolf-Rayet companion and the structure of the wind from the same \citep{vanKerkwijk1996, vanKerkwijk1993}, obtaining the actual mass of the stellar systems is not yet feasible for this source. \par
Nevertheless, it is quite likely that the physical mechanism of mass loss in the system is the cause of the fluctuation of the period change, resulting in the residual change in $\dot{P}$ over and above the secular evolution. The angular momentum can  be closely dependent on the mass loss in the system, therefore the amount of mass ejected from the companion, the fraction of mass accumulated by the accreting compact object and the mass thrown out in the form of jets can all have strong bearings on the angular momentum and result in the fluctuations in the binary modulation of the system. Following the recipe of \cite{Tout1991} the period evolution may be given by 
\begin{equation}
\label{eq:arr_time_2}
    \frac{\dot{P}}{P} = -\frac{2(\dot{m_1}+\dot{m_2})}{(m_1+m_2)} - \frac{3\dot{m_1}(m_2-m_1)}{m_1m_2} + constant ,
\end{equation} \par
where $m_1$ is the mass of the compact object, $m_2$ is the mass of the companion Wolf-Rayet star, $\dot{m_2}$ is the rate of mass change in Wolf-Rayet star ($\dot{m_2}<0$: mass is lost via wind) and $\dot{m_1}$ is the rate of mass change of compact object ($\dot{m_1}>0$: fraction of the wind from companion is accreted, which is perhaps directly related to soft X-ray emission, i.e. higher soft X-ray emission implies higher value of $\dot{m_1}$, because increased accreted mass will cause corresponding incremental change in the angular momentum). 

Assuming the mass gained by the compact object to be a small fraction of the mass lost by the WR star, one may write $\dot{m_1}=-x\dot{m_2}$ where $x$ denotes the fraction of mass deposited on $m_1$, then the value of $x$ can be found from the following relation 
\begin{equation}
\label{eq:arr_time_3}
   x = \frac{1}{\frac{2\dot{m_2}}{m_1+m_2} + \frac{3(m_2-m_1)\dot{m_2}}{m_1m_2}}\left(\frac{\dot{P}}{P}-\frac{2\dot{m_2}}{m_1+m_2}\right) .
\end{equation} \par

Following \cite{Zdziarski2013},  one may assume, as a first approximation, the range of the masses to be $m_1 \approx 1.4 - 3.8 M_\odot$ and $m_2 \approx 8.4 - 11.8 M_\odot$. Similarly the lowest limit of mass loss may be assumed to be $\dot{m_2} \approx 4\times 10^{-5} M_{\odot} / year$ \citep{vanKerkwijk1993,vanKerkwijk1996}. Using these values, one may obtain the upper limit of mass deposited on $m_1$ to be $x=0.23$ and the lower limit to be $x=0.09$ by taking the secular orbital evolution into account. This shows that despite the large amount of the mass that is expected to be lost due to the strong wind, the compact object accretes a significant fraction of the mass despite the presence of the steady jet which throws an unknown fraction of the accreted mass out of the system. A more precise calculation would require a more accurate estimate of the radiative efficiency of the accretion and ejection process. This, in turn, would yield a method of obtaining the mass of the compact object from the X-ray binary modulation, provided the total mass lost from the companion WR star is accurately estimated from the infra-red observations and the total mass lost, via the jets, is obtained from the radio emission.

The physical origin of short period changes has not been explored for this source. The variation of the $\dot{P}$ with the long-term average of soft X-ray flux (Figure \ref{fig:lc10all}) provides a tantalizing opportunity to study the effect of accretion and ejection on the angular momentum of the system.

There is a possibility that the residual period change may arise due to a mild apsidal motion not discovered before.  In general, a proper conclusion regarding the apsidal motion can be reached by simultaneously analysing the arrival time of both primary minima/eclipse (reported above) as well as the secondary minima/eclipse, as the apsidal motion will cause a change in the arrival time of the two minima/eclipses to be exactly out of phase. The binary modulation template of Cyg~X-3 has a smooth maxima without any secondary minima, because although the different light curves do show a two-peaked maxima with a secondary minima in between, the nature of the peaks of the maxima are too varied and diverse to be incorporated in the template, while the monitoring data of the ASM, MAXI or BAT are  too sparse to detect it significantly. Hence, the analysis of the secondary minima, which may be accomplished by analyzing many pointed observations on a semi-continuous basis, might require some future dedicated observations.

The broad hump in the residuals   (see Figure \ref{fig:lc10all}, MJD 52000-55000) may be explained by the factor $\dot{m_1}$ in equation \ref{eq:arr_time_2}. When the soft X-ray emission is dominant and high, with a perhaps higher accretion of mass (effectively angular momentum) from the companion Wolf-Rayet star to the compact object, the $\dot{m_1}$ is high and hence the total $\dot{P}$ decreases. When the soft X-ray emission is low, it indicates lower mass accreted from the companion Wolf-Rayet star to the compact object. This causes the $\dot{m_1}$ value to fall lower resulting in overall $\dot{P}$ to rise above the secular value. Thus for MJD 52000-53500 the $\dot{P}$ rises with fall in soft X-ray emission and hence the drop in the values of $\dot{m_1}$. Around MJD 53500 the state changes to very high values of soft X-rays increasing the value of $\dot{m_1}$ resulting in the lowering of $\dot{P}$ towards the secular value. The mass lost due to outflows and jets also will have an impact on the total angular momentum and hence the overall $\dot{P}$.

\section*{Acknowledgments}
This publication uses the data from the \emph{AstroSat} mission of the Indian Space Research Organisation (ISRO), archived at the Indian Space Science Data Centre (ISSDC). 
The contributions of the technical teams of CZTI and SXT have been vital to the research and thus are gratefully acknowledged.
This research also has made use of data obtained through the High Energy Astrophysics Science Archive Research Center Online Service, provided by the NASA/Goddard Space Flight Center. 
\bibliography{ref.bib}
\end{document}